\documentclass[a4paper]{article}

\usepackage[dvipdfmx,hiresbb]{graphicx}
\usepackage{type1cm}
\usepackage{multirow}
\usepackage{lscape}
\usepackage{amsmath}
\usepackage{framed}
\usepackage{setspace}
\usepackage{float}
\usepackage[title]{appendix}
\usepackage{float}

\usepackage{mathtools}

\usepackage{amsmath,amssymb}

\usepackage{changepage}

\usepackage[utf8x]{inputenc}

\usepackage{textcomp,marvosym}

\usepackage{cite}

\usepackage{nameref,hyperref}

\usepackage[right]{lineno}

\usepackage{microtype}
\DisableLigatures[f]{encoding = *, family = * }

\usepackage[table]{xcolor}

\usepackage{array}

\begin{document}
\fontsize{10.5pt}{18pt}\selectfont

\fontsize{13pt}{18pt}\selectfont
\begin{center}
{\bf The propagation of the economic impact through supply chains:\\ The case of a mega-city lockdown against the spread of COVID-19
}

\vspace{2ex}

\fontsize{11pt}{18pt}\selectfont
Hiroyasu Inoue\footnote{Graduate School of Simulation Studies, University of Hyogo, inoue@sim.u-hyogo.ac.jp.} and
Yasuyuki Todo\footnote{Graduate School of Economics, Waseda University, yastodo@waseda.jp.}
\vspace{3ex}

\fontsize{10.5pt}{18pt}\selectfont
Abstract
\end{center}

\vspace{1ex}
\fontsize{10pt}{12pt}\selectfont

\noindent This study quantifies the economic effect of a possible lockdown of Tokyo to prevent spread of COVID-19. The negative effect of the lockdown may propagate to other regions through supply chains because of shortage of supply and demand. Applying an agent-based model to the actual supply chains of nearly 1.6 million firms in Japan, we simulate what would happen to production activities outside Tokyo when production activities that are not essential to citizens' survival in Tokyo were shut down for a certain period. We find that when Tokyo is locked down for a month, the indirect effect on other regions would be twice as large as the direct effect on Tokyo, leading to a total production loss of 27 trillion yen in Japan, or 5.3\% of its annual GDP. Although the production shut down in Tokyo accounts for 21\% of the total production in Japan, the lockdown would result in a reduction of the daily production in Japan by 86\% in a month.

\noindent {\it Keywords}: COVID-19; Lockdown; Supply-chain; Simulation; Propagation




\fontsize{10pt}{10pt}\selectfont


\section{Introduction}
COVID-19, the novel coronavirus disease, has been spreading all over the world. As of March 31, 2020, the total number of confirmed cases of COVID-19 reached 775,306, whereas the total number of deaths was 37,083 \cite{Corona2020}. To prevent the spread of COVID-19, most countries have implemented unprecedentedly stringent restrictions, such as shutdown of national borders, limits on public gatherings, and closures of schools, shops, and restaurants. 

In some cases, cities and regions are locked down. For example, Wuhan, the epicentre of the novel coronavirus, was locked down from January 23 to March 27, 2020, shutting down all public transports and all companies not essential to citizens' survival including manufacturing plants during most of the period (Reuters, March 11, 2020). Apparently, the lockdown heavily affected Wuhan's economy with a population of 11 million. Moreover, because Wuhan, known as one of China's ``Detroits,'' is a centre of the automobile industry and supplying parts and components of automobiles to domestic and foreign plants, the effect of the lockdown propagated to other regions of China and other countries through supply chains. For example, Honda, a Japanese automobile manufacturer that operates plants in Wuhan reduced production of automobiles in Japan due to lack of supplies of parts from China early March 2020 (Nikkei Newspaper, March 2, 2020).     

Recently, many studies have empirically confirmed that economic shocks propagate across regions and countries through supply chains \cite{Barrot2016, Boehm2019, Carvalho2016, Inoue2019, Kashiwagi2018}, mostly using natural disasters as sources of shocks. For example, Inoue and Todo \cite{Inoue2019} employ the actual supply chains of approximately one million firms in Japan and the production trajectory after the Great East Japan earthquake in 2011 to calibrate an agent-based model with firm-to-firm supply chains. They find that while the production loss because of the direct effect of the earthquake and subsequent tsunamis was approximately 100 billion yen, or 0.02\% of gross national products (GDP), the production loss because of supply chain disruptions in areas that were not directly hit by the earthquake or tsunamis was 11 trillion yen, or 2.3\% of GDP. Their result indicates that the propagation effect of an economic shock through supply chains can be substantially larger than its direct effect. 

Inoue and Todo \cite{Inoue2019} also show that complex network characteristics of supply chains, such as scale-free properties and complex loops, aggravate the propagation effect. Without any network complexity, i.e., if they assume no firm-level inter-linkages but only inter-industry linkages or assume a randomly determined network with no complexity, they find the propagation effect is quite small. These results are consistent with recent findings in the network science literature that the structure of networks significantly influences diffusion \cite{Watts98, Burt04, Centola10, Newman10, Barabasi16, Watts02},

Therefore, when a large industrial city connected with other regions and countries through supply chains in a complex manner is locked down to prevent the spread of COVID-19, the economic effect is most likely to propagate across regions and countries. To confirm this conjecture, we utilise the framework of Inoue and Todo \cite{Inoue2019} and quantify the economic effect of a lockdown of Tokyo on other regions. Tokyo is an appropriate case for the purpose of this study, because it is one of the largest cities in the world and a hub in global supply chains, and because recently policymakers including the governor of Tokyo have mentioned a possibility of a lockdown of Tokyo. Specifically, applying the agent-based model developed in Inoue and Todo \cite{Inoue2019} to the actual supply chains in Japan, we simulate what would happen to production activities outside Tokyo when Tokyo were locked down, or non-essential production activities in Tokyo were shut down for a certain period.   

Several studies have estimated economic impacts of the spread of COVID-19. For example, Organisation for Economic Co-operation and Development (OECD) predicts in early March, 2020 that if outbreaks of COVID-19 spread widely in Asia and advanced countries in the northern hemisphere, the growth rate of real GDP in the world in 2020 would be 1.4\%, which is 1.5\% points lower than its estimate before the spread of COVID-19 \cite{OECD2020}. The estimation of McKibbin and Fernando \cite{McKibbin2020} indicates that in their worst scenario where all countries are hit, the spread of COVID-19 would reduce GDP of China, Japan, the United Kingdom, and the United States by 6.2\%, 9.9\%, 6.0\%, and 8.4\%, respectively. However, these studies rely on either a macroeconomic econometric model at the country level \cite{OECD2020} or a general equilibrium model assuming international and inter-sectoral input-output linkages \cite{McKibbin2020} and thus do not incorporate complex inter-firm linkages. As a result, the estimates of the previous studies may be largely undervalued, as suggested by the finding of Inoue and Todo \cite{Inoue2019}. Therefore, this study attempts to quantify the economic effect of COVID-19 that takes into account propagation of the effect across regions through inter-firm supply chains for the first time in the literature. 

\section{Data}

The data used in this study are taken from the Company Information Database and Company Linkage Database collected by Tokyo Shoko Research (TSR), one of the largest credit research companies in Japan. Because of the data availability, we utilise data for firm attributes and supply chains in 2016 and the IO table in 2015. The former includes information about attributes of each firm, including the location, the industry, sales, and the number of employees, whereas the latter includes major customers and suppliers of each firm. The number of firms in the data is 1,668,567, and the number of supply-chain links is 5,943,073. That is, our data identify major supply chains of most firms in Japan, although they lack information about supply-chain links with foreign entities. Because the transaction value of each supply-chain tie is not available in the data, we estimate sales from a particular supplier to each of its customers and consumers using the total sales of the supplier and its customers and the input-output (IO) table for Japan in 2015. In this estimation process, we have to drop firms without any sales information. Accordingly, the number of firms in our further analysis is 966,627, and the number of links is 3,544,343. Although firms in the TSR data are classified into 1,460 industries according to the Japan Standard Industrial Classification, we simplify them into 187 industries classified in the IO table. Appendix \ref{ch:appdata} provides details of the data construction process.

In the supply-chain data described above, the degree, or the number of links, of firms follows a power-law distribution (Figure \ref{fig:deg}), as often found in the literature \cite{Barabasi16}. The average of the path length between firms, or the number of steps between them through supply chains, is 4.8. This small average path length indicates that the supply chains have a small-world property: i.e., firms are indirectly connected closely through supply chains. Therefore, we would predict that economic shocks quickly propagate through the supply chains. Using the same data set, previous studies \cite{Fujiwara10, Inoue2019} find that 46-48\% of firms are included in the giant strongly connected component (GSCC) in which all firms are indirectly connected to each other through supply chains. The large size of the GSCC prominently shows that the network has numerous cycles and the complex nature, which is unlike the common image of a layered supply-chain structure.

\section{Method}

\subsection{Model}
\label{ch:model}

Our simulation employs the dynamic agent-based model of Inoue and Todo \cite{Inoue2019}, an extension of the model of Hallegatte \cite{Hallegatte08}, that assumes supply chains at the firm level. In the model, each firm utilises inputs purchased from other firms to produce an output and sells it to other firms and consumers. Supply chains are pre-determined and do not change over time in the following two respects. First, each firm utilises a firm-specific set of input varieties and does not change the input set over time. The variety of input is determined by the industry of the producer, and hence firms in a particular industry are assumed to produce the same output. Second, each firm is linked with fixed suppliers and customers and cannot be linked with any new one over time. Further, we assume that each firm keeps inventories of each input at a level randomly determined from the Poisson distribution. Following Inoue and Todo \cite{Inoue2019} where parameter values are calibrated from the case of the Great East Japan earthquake, we assume that firms target to keep inventories for nine days of production on average.

When a lockdown directly or indirectly causes a reduction in production of particular firms, the supply of products of these firms to their customer firms declined. Then, one way to keep the current level of production of the customers is to use their inventories of inputs. Alternatively, the customers can procure the input from their other suppliers in the same industry already connected prior to the lockdown if these suppliers have additional production capacity. If the inventories and inputs from substitute suppliers are insufficient, the customers have to shrink their production because of shortage of inputs. In addition, suppliers of the firms directly affected by the lockdown may have to reduce production because of the reduction of demand from the affected customers. Accordingly, the economic shock propagates both downstream and upstream through supply chains. Appendix \ref{ch:appmodel} provides details of the model. 

\subsection{Simulation Procedure}

In the simulation, we assume that all production activities not essential to citizens' survival (hereafter, referred to as non-essential production activities) in the central part of Tokyo (23 wards, hereafter simply referred to as Tokyo) are shut down for either one day, one week, two weeks, one month, or two months. Essential production activities are defined as those in the wholesale, retail, utility, transport, storage, communication, healthcare, and welfare sectors\footnote{According to the industry classification used in the 2015 national IO table of Japan, these are coded 4611, 4621, 4622, 4711, 4811, 5111, 5112, 5711, 5712, 5721, 5722, 5741, 5742, 5743, 5761, 5771, 5781, 5789, 5791, 5911, 5921, 5931, 5941, 5951, 6411, 6421, 6431, and 6441.}. After the lockdown period, all sectors immediately resume production at the same level as in the pre-lockdown period. Because the inventory target of each firm is randomly sampled from the Poisson distribution (Section \ref{ch:model}), we run five simulations with different sets of the inventory targets of firms for each lockdown duration and average over the five sets of results. 

\section{Result}

\subsection{Benchmark Result}

When Tokyo is locked down, value added production of Tokyo immediately becomes almost zero. Because the daily production of non-essential sectors in Tokyo is estimated to be 309 billion yen, or approximately 2.9 billion US dollars, the total direct loss of production in Tokyo because of the lockdown is 309 billion yen multiplied by the number of days during the lockdown period. Table \ref{tbl:loss} shows the direct production loss in Tokyo for each case in the second column and additionally the production loss outside Tokyo because of the propagation effect through supply chains in the third column. These results are the averages of the simulations.

\begin{figure}[tbh]
\centering
\includegraphics[width=0.6\linewidth]{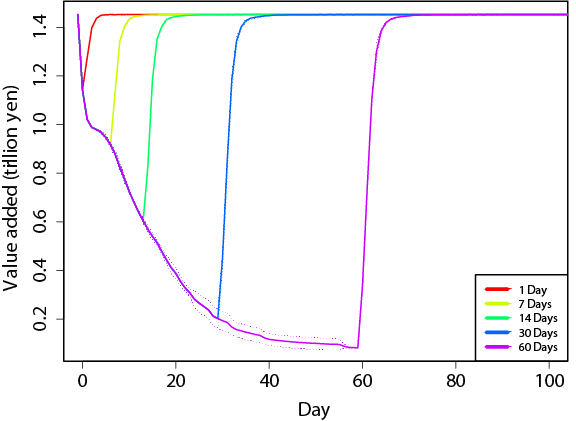}
\caption{The dynamics of daily value added in Japan after the lockdowns of Tokyo. Each line shows the average of five simulations assuming different inventory sizes. The dotted lines show the standard deviations. This figure shows simulation results assuming shutdown of all non-essential production activities.}
\label{fig:effect2}
\end{figure}

The results indicate that when Tokyo is locked down for only one day, the production loss outside Tokyo, though not locked down, is already 252 billion yen, 82\% of the production loss in Tokyo. When the lockdown continues for a month, the indirect effect on other regions is twice as large as the direct effect on Tokyo, and the estimated total production loss is 27.8 trillion yen, or 5.25\% of the annual GDP. In Figure \ref{fig:effect2}, each line shows dynamics of total daily value added in Japan in each case assuming different lockdown duration. It is shown that when the lockdown continues for a month, daily value added production of Japan becomes only approximately 1/7 of that before the lockdown. This implies that even when the initial production loss in Tokyo is small, its propagation effect on other regions can become large as the lockdown prolongs.

\begin{figure}[tbh]
\centering
\includegraphics[width=1\linewidth]{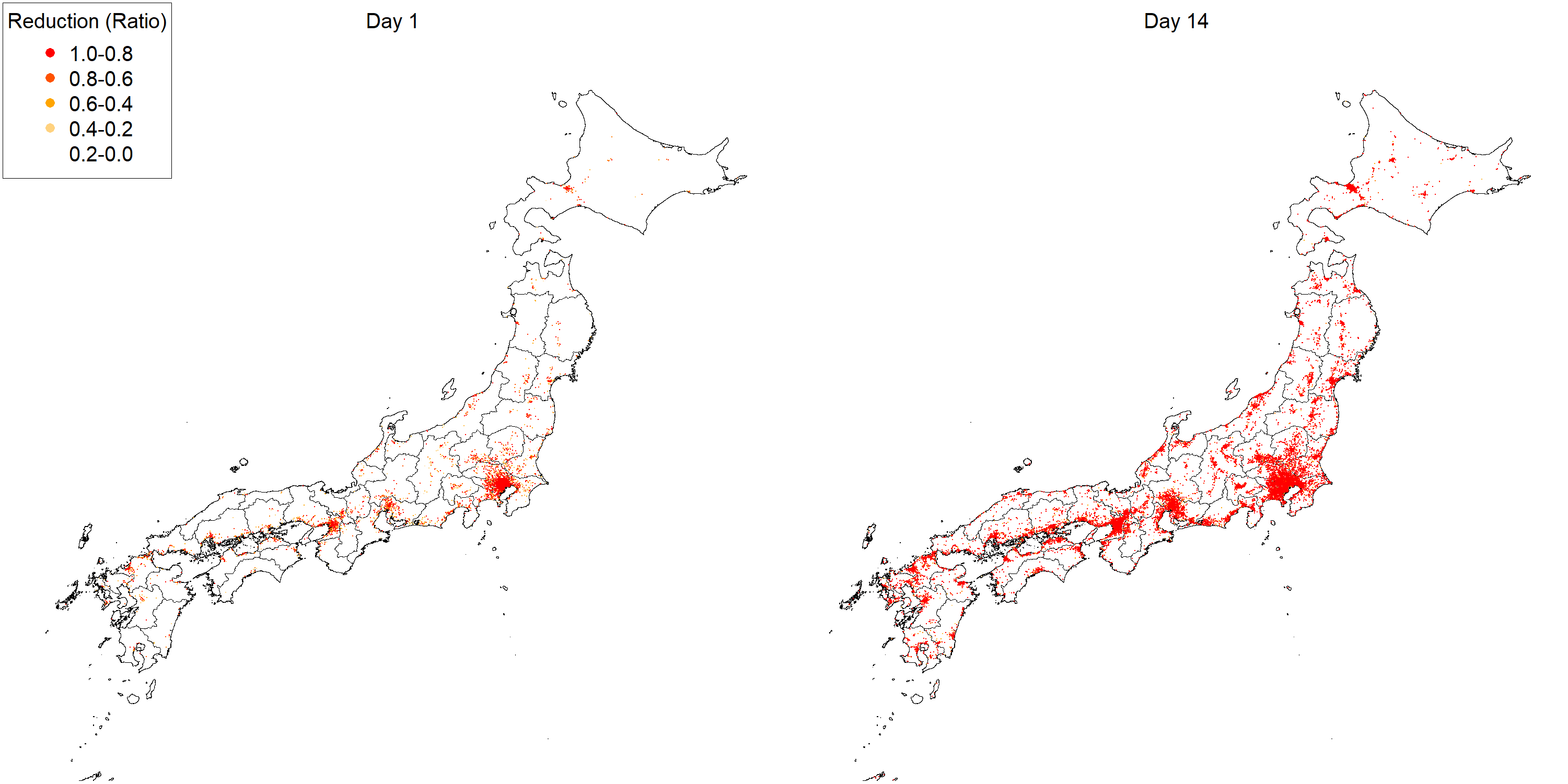}
\caption{Temporal and geographical visualisations of the reduction of the production. The left and right panels show the first day and the 2 weeks after the lockdown respectively.
The red and orange dots indicate firms whose actual production is substantially and moderately, respectively, smaller than their capacity before a lockdown.
Since there are more than 1 million dots in the original data, we randomly choose 100,000 firms to draw.}
\label{fig:map}
\end{figure}

Figure \ref{fig:map} shows temporal and geographical visualisations of the simulation of a lockdown. The red dots indicate firms whose production is less than or equal to 20\% of their capacity, whereas the light red and orange dots show firms with a more moderate decline in production. The left figure illustrates that a non-negligible number of firms distant from Tokyo are already affected on the first day of the lockdown. Two weeks later, affected firms spread all over the country, as shown in the right figure. These visualisations support that the indirect effect propagates geographically as the lockdown is prolonged.

\begin{table}[tbh]
\centering
\caption{The loss of value added because of a Tokyo lockdown. This table shows the results from the simulations assuming shutdown of all non-essential production activities. These results are based on the average of the simulations. (Unit: trillion yen)}
\begin{tabular}{lrrr}
\hline
 & \multicolumn{1}{l}{Direct effect} & \multicolumn{1}{l}{Indirect effect} & \multicolumn{1}{l}{Total effect} \\
 & \multicolumn{1}{l}{on Tokyo} & \multicolumn{1}{l}{on other regions} & \multicolumn{1}{l}{(\% of GDP)} \\
 & & \multicolumn{1}{l}{in Japan} & \\
\hline
1 day & 0.309 & 0.252 & 0.561 (0.106) \\
1 week & 2.17 & 1.56 & 3.72 (0.720) \\
2 weeks & 4.33 & 5.01 & 9.34 (1.76) \\
1 month & 9.28 & 18.5 & 27.8 (5.25) \\
2 months & 18.6 & 50.0 & 68.2 (12.9) \\
\hline
\end{tabular}
\label{tbl:loss}
\end{table}

\subsection{Alternative Specifications}

In addition to the benchmark simulations above, we experiment with two alternative sets of simulations. First, we assume that all production activities including essential activities are shut down. Then, the daily production loss in Tokyo is 471 billion yen, 52\% larger than that in the benchmark simulations (309 billion yen). However, we find that the total production loss in Japan from a lockdown of Tokyo for a month is 32.0 trillion yen, only 15\% larger than that in the benchmark (27.8 trillion yen). (See Appendix \ref{ch:appalt} for the detail.)

Second, we assume that industrial demand is prioritised over consumer demand so that production activities outside Tokyo would be less affected. For example, computers can be used both by customer firms for production and citizens for consumption. In the benchmark simulation, when the output of a product is not sufficient because of a lockdown, we assume that the limited output is rationed to customer firms and consumers based on their relative demand prior to the lockdown. However, in this alternative simulation, customer firms are prioritised to maximise production in downstream firms. (See Appendix \ref{ch:appmodel} for the detail.) Then, we find that a one-month lockdown results in a production loss in Japan of 27.0 trillion yen. 
Because the production loss does not substantially change from the benchmark result (27.8 trillion yen), we conclude that industry prioritisation is not much effective to alleviate the propagation effect of a lockdown.

\section{Discussion and Conclusion}

The simulation results clearly show that the effect of a lockdown of Tokyo quickly propagates to other regions outside Tokyo, leading to a substantial effect on the entire Japanese economy. Although the production of non-essential sectors in Tokyo accounts for 21.3\% of the total production in Japan, a lockdown of Tokyo for a month would result in a reduction of the daily production in Japan by 86.1\% or 1.25 trillion yen.

In addition, the effect on other regions becomes progressively larger, as the duration of the lockdown becomes longer: When the duration becomes twice, the production loss becomes more than twice. In the case of a lockdown for one day, the total loss of value added outside Tokyo is 82\% of the loss in Tokyo. However, when the lockdown continues for a month, the loss outside Tokyo is twice as large as the loss in Tokyo. This implies that the effect of a longer lockdown can reach firms that are ``further'' from Tokyo along supply chains.  

To alleviate the propagation effect through supply chains, one could limit production activities shut down or prioritise producers' use of goods and servies over consumers' use. However, our results indicate that these measures would not work well, particularly when the duration of the lockdown is long. 

Our analysis provides several policy implications. First, because the overall effect of a lockdown of a major city on the entire economy is extremely large when we take into account its propagation effect through supply chains, we should consider lockdowns as the last resort. Rather, we should prevent the spread of COVID-19 earlier using other means and avoid any lockdown of a mega city. Second, because the total effect of a lockdown progressively increases with its duration, a mega-city lockdown, even if it cannot be avoided, should be as short as possible. Policymakers should be aware that policies to alleviate the propagation effect may not work when the lockdown duration is long.  

Several caveats of this study should be mentioned. First, we assume that firms cannot find any new supplier when supplies from their suppliers in Tokyo are disrupted, although they can request their existing suppliers outside Tokyo to supply more. This assumption may be too strong in practice, leading to an overestimation of the propagation effect. However, because we particularly examine a short-term lockdown for at most two months, the possible overestimation can be minimal as finding new suppliers in the short period of time is not easy.  

Second, the TSR data reports only the location of the headquarter of each firm, not the location of its branches. Because headquarters of firms concentrate in Tokyo, production activities in Tokyo are most likely to be overvalued in our analysis. Therefore, the direct effect of a lockdown of Tokyo may be overestimated while its propagation effect on other regions may be underestimated. Because we still found a large propagation effect despite of this possible underestimation, our key conclusion should remain the same.

Third, because of the data limitation, we cannot estimate the propagation effect of a lockdown of Tokyo on foreign economies outside Japan. Kashiwagi et al. \cite{Kashiwagi2018} find no international propagation effect in the case of Hurricane Sandy in the United States, suggesting that substitution for damaged suppliers can alleviate propagation. However, in the case of the spread of COVID-19, because all industrial countries are affected, input substitution across countries is quite difficult. Therefore, we would expect international propagation of the economic effect of a lockdown of a city amid the spread of COVID-19, but quantifying this propagation is beyond the scope of this study. 

Finally, we should emphasise that this study focuses on the economic effect of a lockdown of a major industrial city, rather than the overall economic effect of COVID-19. In practice, COVID-19 should affect economies outside a mega city not only through propagation of the effect of its lockdown but also directly through other restrictions in the region and indirectly through propagation of the effect of lockdowns of foreign cities. Because these effects are not included in the simulation of this study, the total economic effect of COVID-19 can be considerably larger than estimated here.  

\bibliographystyle{unsrt}
\bibliography{reference}

\begin{appendices}
\renewcommand\thefigure{\thesection.\arabic{figure}} 
\renewcommand\thetable{\thesection.\arabic{table}} 

\section{Degree distribution of supply-chain network}

\setcounter{figure}{0} 
\setcounter{table}{0} 

\begin{figure}[tbh]
\centering
\includegraphics[width=0.4\linewidth]{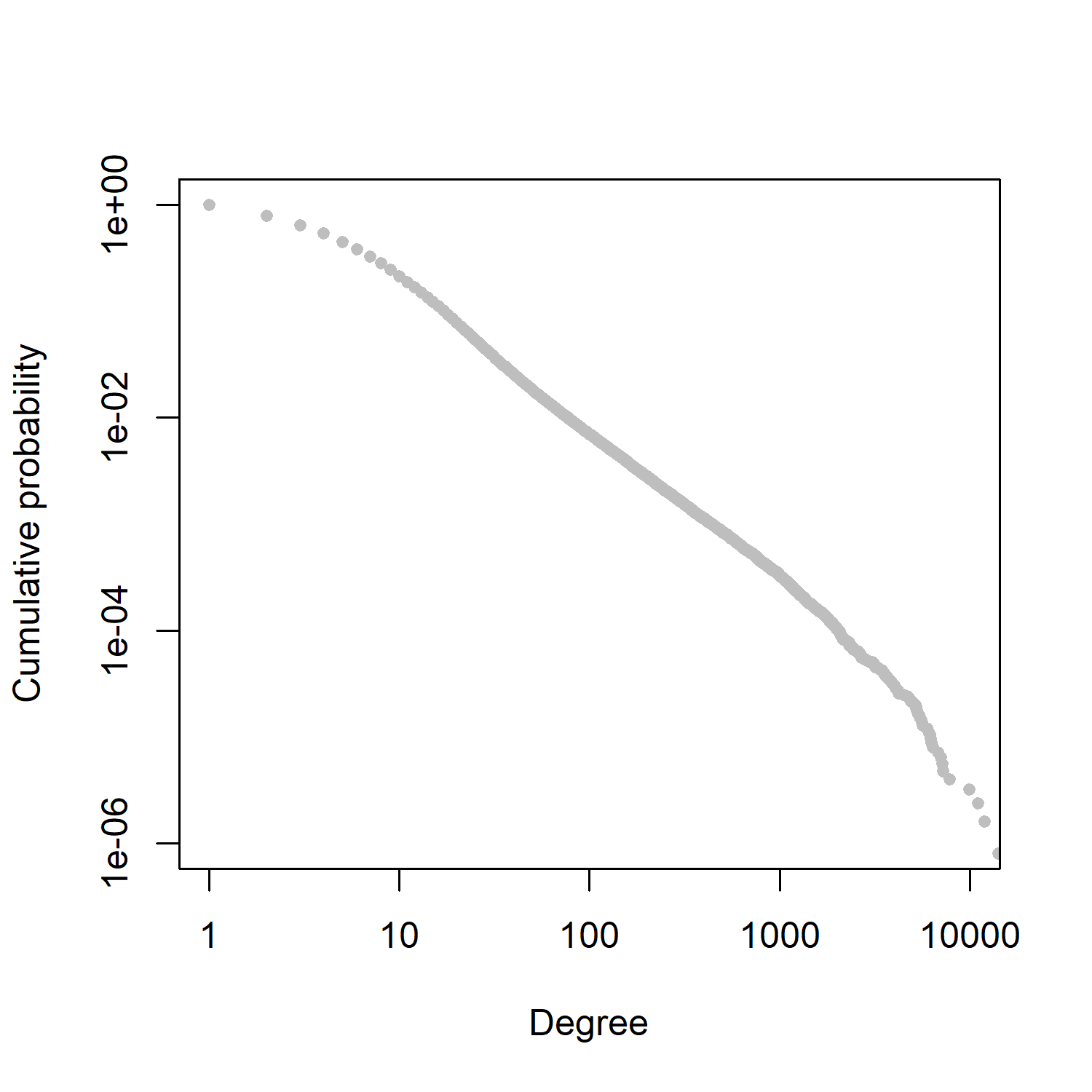}
\caption{Degree distribution of supply chains in Japan.}
\label{fig:deg}
\end{figure}

\section{Data}
\label{ch:appdata}
\setcounter{figure}{0} 
\setcounter{table}{0} 

In the TSR data, the maximum number of suppliers and customers reported by each firm is 24. However, we can capture more than 24 by looking at the supplier--customer relations from the opposite direction. Because the TSR data include the address of the headquarters of each firm, we can identify the longitude and latitude of each headquarters by using the geocoding service provided by the Center for Spatial Information Science at the University of Tokyo.

We estimate the value of each transaction between two firms in two steps. First, we divide each supplier's sales into its customers in proportion to the sales of customers, defining a tentative sales value. Second, we employ the IO table for Japan in 2015 \cite{METI15} to transform these tentative values into more realistic ones. Specifically, we aggregate the tentative values at the firm--pair level to obtain the total sales for each pair of sectors. We then divide the total sales for each sector pair by the transaction values for the corresponding pair in the IO tables. The ratio is then used to estimate the transaction values between firms. The final consumption of each sector is allocated to all firms in the sector, using their sales as weights. 

\section{Model}
\label{ch:appmodel}
\setcounter{figure}{0} 
\setcounter{table}{0} 

\begin{figure}[thbp]
\centering
\includegraphics[width=0.5\linewidth]{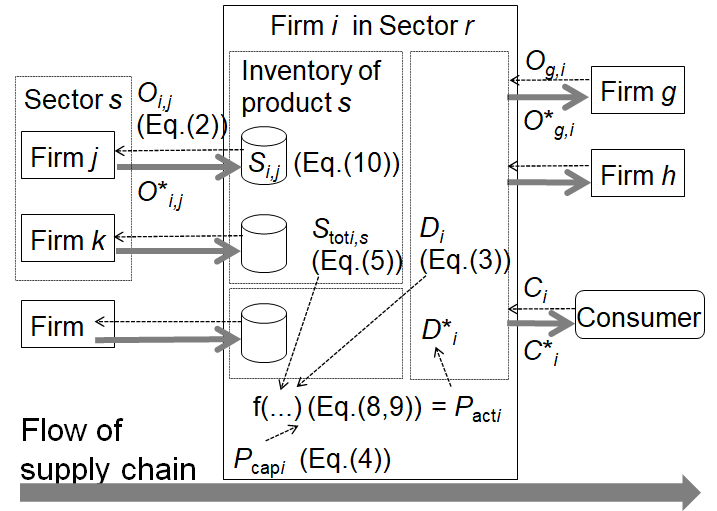}
\caption{Overview of the agent-based model. Products flow from left to right, whereas orders flow in the opposite direction. The equation numbers correspond to those in Appendix \ref{ch:model}.}
\label{fig:model}
\end{figure}

We rely on the model of Inoue and Todo \cite{Inoue2019}, an extension of existing agent-based models used to examine the propagation of shocks by natural disasters through supply chains of Hallegatte \cite{Hallegatte08}. Each firm uses a variety of intermediates as inputs and delivers a sector--specific product to other firms and the final consumers. Firms have an inventory of intermediates to deal with possible supply shortages. Figure \ref{fig:model} provides an overview of the model, showing the flows of products to and from firm $i$ in sector $r$.

In the initial stage before an economic shock, the daily trade volume from supplier $j$ to customer $i$ is denoted by $A_{i,j}$, whereas the daily trade volume from firm $i$ to the final consumers is denoted as $C_i$. Then, the initial production of firm $i$ in a day is given by 
\begin{equation}
P_{\mbox{ini}i}=\Sigma_j{A_{j,i}}+C_i.
\label{eq:p}
\end{equation}
On day $t$ after the initial stage, the previous day's demand for firm $i$'s product is $D_i^* (t-1)$. The firm thus make orders to each supplier $j$
so that the amount of its product of supplier $j$ can meet this demand, $A_{i,j}{D_i^*(t-1)}/{P_{\mbox{ini}i}}$. We assume that firm $i$ has an inventory of the intermediate goods produced by firm $j$ on day $t$, $S_{i,j} (t)$, and aims to restore this inventory to a level equal to a given number of days $n_i$ of the utilization of product of supplier $j$. The constant $n_i$ is assumed to be Poisson distributed, where its mean is $n$, which is a parameter. That is, when the actual inventory is smaller than its target, firm $i$ increases its inventory gradually by $1/\tau$ of the gap, so that it reaches the target in $\tau$ days, where $\tau$ is assumed to be six to follow the original model \cite{Hallegatte08}. Therefore, the order from firm $i$ to its supplier $j$ on day $t$, denoted as $O_{i,j}(t)$, is given by 
\begin{equation}
O_{i,j}(t)=A_{i,j}\frac{D_i^*(t-1)}{P_{\mbox{ini}i}}+\frac{1}{\tau}\left[n_i A_{i,j}-S_{i,j}(t)\right],
\label{eq:o}
\end{equation}
where the inventory gap is in brackets. Accordingly, total demand for the product of supplier $i$ on day $t$, $D_i(t)$, is given by the sum of final demand from final consumers and total orders from customers:
\begin{equation}
D_i(t)=\Sigma_jO_{j,i}(t)+C_i,
\end{equation}

Now, suppose that an economic shock hits the economy on day 0 and that firm $i$ is directly damaged. Subsequently, the proportion $\delta_{i}(t)$ of the production capital of firm $i$ is malfunctioning, although $\delta_{i}(t)$ decreases over time because of the recovery effort, as we explain in the following paragraph. Hence, the production capacity of firm $i$, defined as its maximum production assuming no supply shortages, $P_{\mbox{cap}i}(t)$, is given by
\begin{equation}
P_{\mbox{cap}i}(t)=P_{\mbox{ini}i}(1-\delta_i(t)).
\end{equation}
The production of firm $i$ might also be limited by the shortage of supplies on day 0. Because we assume that firms in the same sector produce the same product, shortage of supplies suffered by firm $j$ in sector $s$ can be compensated for by supplies from firm $k$ in the same sector. Firms cannot substitute new suppliers for damaged ones after the disaster, as we assume fixed supply chains. Thus, the total inventory of the products delivered by firms in sector $s$ in firm $i$ on day $t$ is
\begin{equation}
S_{\mbox{tot}i,s}(t)=\Sigma_{j\in s}S_{i,j}(t).
\end{equation}
The initial consumption of products in sector $s$ at firm $i$ before the disaster is also defined for convenience:
\begin{equation}
A_{\mbox{tot}i,s}=\Sigma_{j\in s}A_{i,j}.
\end{equation}
The maximum possible production of firm $i$ limited by the inventory of product of sector $s$ on day $t$, $P_{\mbox{pro}i,s}(t)$, is given by
\begin{equation}
P_{\mbox{pro}i,s}(t)=\frac{S_{\mbox{tot}i,s}(t)}{A_{\mbox{tot}i,s}}P_{\mbox{ini}i}.
\end{equation}
Then, we can determine the maximum production of firm $i$ on day $t$, considering its production capacity, $P_{\mbox{cap}i}(t)$, and its production constraints due to the shortage of supplies, $P_{\mbox{pro}i,s}(t)$:
\begin{equation}
P_{\mbox{max}i}(t)=\mbox{Min}\left(P_{\mbox{cap}i}(t), \mbox{Min}_{s}(P_{\mbox{pro}i,s}(t))\right). \label{eq:pconst}
\end{equation}
Therefore, the actual production of firm $i$ on day $t$ is given by
\begin{equation}
P_{\mbox{act}i}(t)=\mbox{Min}\left(P_{\mbox{max}i}(t), D_i(t)\right).
\label{eq:act}
\end{equation}

When demand for a firm is greater than its production capacity, the firm cannot completely satisfy its demand, as is denoted by Equation (9). In this case, firms should ration their production to their customers. We propose a rationing policy in which customers and final consumers are prioritized according to their amount of order after the economic shock to their initial order, rather than they are treated equally as in the previous work\cite{Hallegatte08}. Suppose that firm $i$ has customers $j$ and a final consumer. Then the ratio of the order from customers $j$ and the final consumer after the shock to the one before the shock denoted as $O^{rel}_{j,i}$ and $O^{rel}_{c}$, respectively, are determined by the following steps, where $O^{sub}_{j,i}$ and $O^{sub}_{c}$ are temporal variables to calculate the realized order and set to be zero initially. 
\begin{enumerate}
\item Get the remaining production $r$ of firm $i$
\item Calculate $O^{rel}_{\mbox{min}}=\mbox{Min}(O^{rel}_{j,i}, O^{rel}_{c})$
\item If $r \leq (\sum_{j}{O^{rel}_{\mbox{min}}O_{j,i}}+O^{rel}_{\mbox{min}}C_i)$ then proceed to 8
\item Add $O^{rel}_{\mbox{min}}$ to $O^{sub}_{j,i}$ and $O^{sub}_{c}$ 
\item Subtract $(\sum_{j}{O^{rel}_{\mbox{min}}O_{j,i}}+O^{rel}_{\mbox{min}}C_i)$ from $r$ 
\item Remove the customer or the final consumer that indicated $O^{rel}_{\mbox{min}}$ from the calculation
\item Return to 2
\item Calculate $O^{rea}$ that satisfies $r=(\sum_{j}{O^{rea}O_{j,i}}+O^{rea}C_i)$
\item Get $O_{j,i}^*=O^{rea}O_{j,i}+O^{sub}_{j,i}O_{j,i}$ and $C_i^*=O^{rea}C_{i}+O^{sub}_{c}C_{i}$,
where the realized order from firm $j$ to supplier $i$ is denoted as $O_{j,i}^* (t)$, and the realized order from a final consumer is $C_i^*$
\item Finalize the calculation
\end{enumerate}

The above rationing is used for the benchmark simulations. In the alternative specification with priority of industry, we use the same algorithm but final consumers are excluded from the calculation. Instead, only when all demand of the customer firms are fulfilled, the remaining production after rationing customer firms is assigned to the final consumer.

Under this rationing policy, total realized demand for firm $i$, $D_i^* (t)$, is given by
\begin{equation}
D_i^*(t)=\Sigma_jO_{i,j}^*(t)+C_i^*,
\end{equation}
where the realized order from firm $i$ to supplier $j$ is denoted as $O_{i,j}^*(t)$ and that from the final consumers is $C_i^*$. According to firms' production and procurement activities on day $t$, the inventory of firm $j$'s product in firm $i$ on day $t+1$ is updated to
\begin{equation}
S_{i,j}(t+1)=S_{i,j}(t)+O^{*}_{i,j}(t)-A_{i,j}\frac{P_{\mbox{act}i}(t-1)}{P_{\mbox{ini}i}}.
\end{equation}

\section{Results from Alternative Specifications}
\label{ch:appalt}
\setcounter{figure}{0} 
\setcounter{table}{0} 

\begin{table}[H]
\centering
\caption{The loss of value added because of a Tokyo lockdown on alteranative specifications. (Unit: trillion yen)}
\begin{tabular}{lrrrr}
\hline
 & \multicolumn{1}{l}{Direct effect} & \multicolumn{1}{l}{Indirect effect} & \multicolumn{1}{l}{Total effect} & \multicolumn{1}{l}{Total effect} \\
 & \multicolumn{1}{l}{on Tokyo} & \multicolumn{1}{l}{on other regions} & \multicolumn{1}{l}{(\% of GDP)} & \multicolumn{1}{l}{difference} \\
 & & \multicolumn{1}{l}{in Japan} & & \multicolumn{1}{l}{with benchmark}\\
\hline
\multicolumn{5}{l}{A. All production activities are shut down}\\
1 day & 0.471 & 0.349 & 0.820 (0.155) & 0.259 \\
1 week & 3.30 & 1.58 & 4.88 (0.922) & 1.16 \\
2 weeks & 6.60 & 4.88 & 11.47 (2.17) & 2.13 \\
1 month & 14.1 & 17.8 & 31.96 (6.04) & 4.16 \\
2 months & 28.3 & 44.5 & 72.74 (13.7) & 4.54 \\
\hline
\multicolumn{5}{l}{B. Producers' use is prioritised over final consumers' use}\\
1 day & 0.309 & 0.252 & 0.561 (0.106) & -9.06$\times 10^{-4}$ \\
1 week & 2.17 & 1.53 & 3.69 (0.700) & -3.09$\times 10^{-3}$ \\
2 weeks & 4.33 & 4.87 & 9.20 (1.74) & -2.27 \\
1 month & 9.28 & 17.7 & 27.0 (5.09) & -4.96 \\
2 months & 18.6 & 47.9 & 66.4 (12.5) & -6.34 \\
\hline
\end{tabular}
\label{tbl:loss2}
\end{table}
\end{appendices}

%

\section*{Acknowledgement}
This research was conducted as part of a project entitled ‘Large-scale Simulation and Analysis of Economic Network for Macro Prudential Policy,’ undertaken at the Research Institute of Economy, Trade, and Industry. This research was also supported by MEXT
as Exploratory Challenges on Post-K computer (Studies of Multi-level Spatiotemporal Simulation of Socioeconomic Phenomena, Macroeconomic Simulations). This research used computational resources of the K computer provided by the RIKEN Advanced Institute for Computational Science through the HPCI System Research project (Project ID: hp190148). The authors are grateful for the financial support of JSPS Kakenhi Grant Nos. 18K04615 and 18H03642. The opinions expressed and arguments employed herein do not necessarily reflect those of RIETI, University of Hyogo, Waseda University, or any institution with which the authors are affiliated.

\end{document}